# Energy dissipation processes in solar wind turbulence

Y. Wang[1,2,3,*], F. S. Wei[1], X. S. Feng[1], X. J. Xu[4,5], J. Zhang[3], T. R. Sun[1] and P. B. Zuo[1]

1 SIGMA Weather Group, State Key Laboratory for Space Weather, National Space Science Center,
Chinese Academy of Sciences, Beijing 100190, China
2 Key Laboratory of Geospace Environment, University of Science and Technology of China,
Chinese Academy of Sciences, Hefei 230026, China
3 School of Physics, Astronomy and Computational Sciences, George Mason University, 4400 University Drive,
MSN 3F3, Fairfax, Virginia 22030, USA
4 Space Science Institute, Macau University of Science and Technology, Macao, China
5 Institute of Space Science and Technology, Nanchang University, Nanchang 330031, China
*email: yw@spaceweather.ac.cn

**Abstract**
Turbulence is a chaotic flow regime filled by irregular flows. The dissipation of turbulence is a fundamental problem in the realm of physics. Theoretically, dissipation cannot be ultimately achieved without collisions, and so how turbulent kinetic energy is dissipated in the nearly collisionless solar wind is a challenging problem. Wave particle interactions and magnetic reconnection are two possible dissipation mechanisms, but which mechanism dominates is still a controversial topic. Here we analyze the dissipation region scaling around a solar wind magnetic reconnection region. We find that the magnetic reconnection region shows a unique multifractal scaling in the dissipation range, while the ambient solar wind turbulence reveals a monofractal dissipation process for most of the time. These results provide the first observational evidences for the intermittent multifractal dissipation region scaling around a magnetic reconnection site, and they also have significant implications for the fundamental energy dissipation process.

**Introduction**
Turbulence exists commonly in the universe (Biskamp 2003; Frisch 1995). In hydrodynamics, the turbulent kinetic energy cascades inviscidly from large-scale structures to small-scale structures in the inertial range. This process continues to generate smaller and smaller hierarchical structures until the viscosity dominates in the dissipation range where the turbulent energy is finally dissipated. Similar scenarios can be found in the solar wind where there is full of magnetohydrodynamic turbulence (Biskamp 2003; Falgarone & Passot 2003; Matthaeus & Velli 2011). Observations have shown that the power spectrum of magnetic field in the solar wind near 1 AU has a Kolmogorov-like power law (Kolmogorov 1941) in the inertial range and it begins to reveal a steeper spectrum in the dissipation range (Alexandrova, Lacombe, & Mangeney 2008; Alexandrova et al. 2012; Leamon et al. 1998; Sahraoui et al. 2009). Such a steeper spectrum is an indication of energy dissipation, but the dissipation mechanism in solar wind turbulence is not the same as those in hydrodynamics since the mean free path of particles in the solar wind is on the order of 1 AU so that the classical molecular collisions cannot be counted on. In collisionless plasma, cyclotron damping and Landau damping are able to heat particles. In addition, plasma can also be heated and accelerated near the current sheets especially when magnetic reconnection (MR) occurs (Birn & Priest 2007). After decades of research, it has been found that the collisionless dissipation mechanisms can be broadly grouped into two categories: wave particle interactions and current sheets/magnetic reconnection (Bale et al. 2005; Howes et al. 2011; Karimabadi et al. 2013b; Matthaeus et al. 2003; Osman et al. 2014; Osman et al. 2012b; Pablo, Matthaeus, & Seenu 2004; Parashar & Salem 2013; TenBarge & Howes 2013). However, up to now, debates still exist among these different mechanisms (Parashar & Salem 2013), and it is of prime importance to understand the dominated dissipation mechanism in collisionless magnetohydrodynamics.

Before answering the dissipation question, it should be noted that the reconnection current sheets can be produced by turbulence, and conversely, turbulence can also be generated by the MR processes (Eastwood et al. 2009; Karimabadi et al. 2013a; Liu et al. 2013; Retino et al. 2007; Servidio et al. 2009; Servidio et al. 2010). Therefore, the so-called 'solar wind turbulence' is composed of at least two types of turbulence, the 'new' turbulence generated locally by MR processes and the 'background' solar wind turbulence. Observations (Abraham & Pablo 2011) show that the MR-generated turbulence in the vicinity of a front magnetic cloud boundary layer reveals a Kolmogorov power spectrum in the inertial subrange, and its scaling properties resemble the She–Leveque magnetohydrodynamic model. Analyses of the spectral scalings and spectral break locations to the reconnection outflow in the solar wind (Vörös et al. 2014) also imply that the reconnection outflow can locally generate turbulence and determine the particular local field and plasma conditions, which may be in favor of one dissipation mechanism or another in the turbulent solar wind. Since it is not known whether these disparate types of turbulence undergo the same energy dissipation process, it is necessary to know whether or not there exists any difference between the 'new' MR-generated turbulence and the 'background' solar wind turbulence, and if there is, then what their properties would be. The answer to these questions shall shed light on the dissipative mechanisms.

Several existing studies have shown that turbulence can differ in certain properties, such as the intermittency and scaling properties of the magnetic field. In particular, turbulence tends to be more intermittent when it is associated with the local heating or MR processes (Karimabadi, et al. 2013b; Osman et al. 2011; Osman et al. 2012a; Osman, et al. 2012b; Wan et al. 2012). Moreover, it has also been revealed that turbulence in the fast ambient solar wind has a monoscaling property in the dissipation range (Kiyani et al. 2009), while in the turbulence generated by collisionless MR, the dissipation process has been numerically demonstrated to be multifractal (Leonardis et al. 2013). Therefore, analyzing the properties of the dissipation region scaling in the MR region and comparing their differences with those in the ambient solar wind could be a meaningful way to reveal its dissipative nature. In this article, we find an MR region in the solar wind and compare its turbulent properties with those in the nearby ambient solar wind. It is found that the region around an MR site has a multifractal scaling in the dissipation range that is removed when the surrounding turbulence is included, and so it may have been unintentionally overlooked in previous analyses. These results indicate that the dissipation of magnetic energy in MR is caused by intermittent cascade through multifractal processes which are different from those in the ambient solar wind.

## Observations and analyses

The region where MR occurs is usually referred to as the reconnection diffusion region (Birn & Priest 2007) whose scale is of the order of the ion inertial length ($\sim 10^2$ km in the solar wind). It is very difficult for current spacecraft to resolve such a region in the ~400km/s solar wind. However, the region dominated by MR-generated turbulence is much larger than the ion inertial scale (Eastwood, et al. 2009; Karimabadi, et al. 2013a; Liu, et al. 2013; Servidio, et al. 2009; Servidio, et al. 2010), so it is possible to analyze the dissipation region scaling in the MR region if the trajectory of the spacecraft is sufficiently close the reconnection diffusion region. Therefore, we focus on a dynamic region called magnetic cloud boundary layer (BL) (Wang et al. 2010; Wang et al. 2012; Wei et al. 2006; Wei et al. 2003) to search the MR region. As seen in Fig.1, the BL is located in front of the magnetic cloud (MC) body. The boundary of the BL on the MC side is just the beginning of the MC body, while the other boundary is determined by systematic analyses of the magnetic field and plasma signatures (Wei, et al. 2003). Under the conditions of high magnetic Reynolds numbers in the interplanetary space, previous studies have shown that the BL is a turbulent layer formed by the interaction of the MC and the ambient solar wind. The magnetic field, plasma temperature, and density behavior inside the BL are completely different from those in the ambient solar wind (Abraham & Pablo 2011; Wang, et al. 2012; Wei, et al. 2006). Compared with those in the nearby upstream solar wind, statistical analyses show that the magnitude of the magnetic field (|B|), proton density (N), proton temperature (Tp) and electron temperature (Te) in the BL change by ~-16.4%, +42.9%, +16.6%, and +5.3% respectively. These behaviors are similar to the variations found in the reconnection exhaust (where |B|, N, Te, and Tp change by~-20.1%, +35.8%, +10.6%, and +27.1% respectively) (Gosling et al. 2007; Gosling et al. 2005; Wang, et al. 2012). In addition, it has also been found that the proton and electron flux variations in the BLs resemble those in the reconnection exhaust (Wang, et al. 2012). All of these features demonstrate that the BL displays the same properties as those in the reconnection exhaust and MR processes could prevail in these regions. Therefore, the BL is a suitable place to seek the MR region.

On 1997 April 21, the WIND spacecraft encountered a BL event at $x,y,z$=[225,2,24]Re (Earth radii) in geocentric solar ecliptic (GSE) coordinates (sketched in Fig.1). The detailed solar wind parameters are plotted in Fig.2. The upstream solar wind, BL, and MC body are marked by 'SW', 'BL' and 'MC' respectively. Inside the BL, the average solar wind speed is $V_{sw}$~400km/s, the magnetic field is |B|~3.3nT, the proton density is N~28cm$^{-3}$, the proton temperature is Tp~4.6eV, the Alfvén speed is $V_A$~13km/s, the ion inertial length is $d_i$~43km, the average angle between the magnetic field and the bulk solar wind velocity vectors is $\theta_{VB}$~56°, and the r.m.s. value of |B| is ~0.98. Note that the magnetic field reverses ~140° across the BL and the proton density and temperature both increase inside the BL. Moreover, around 11:55UT when the magnetic field decreases from ~5nT to ~1nT, a pair of ~10km/s jets pointing to the opposite directions are consecutively observed. The speed of these jets is also close to the local Alfvén speed, which is a typical characteristic of reconnection jets. All of these features indicate that the WIND spacecraft detects an MR region and the reconnection site locates around 11:55UT. Then, we pay a great amount of attention to the properties of the dissipation region scaling during 11:50-12:00UT. The magnetic field data are all obtained from the WIND/MFI instrument (Lepping et al. 1995), and the time resolution is 0.092s.

The heating of the solar wind has been indicated to be inhomogeneous, and the dissipation of solar wind turbulence might be highly connected to the intermittent structures such as the current sheets and the probably associated MR processes (Karimabadi, et al. 2013b; Osman, et al. 2011; Osman, et al. 2012a; Osman, et al. 2012b; Wan, et al. 2012). First, we analyze the probability distribution functions (PDFs) and kurtosis $\kappa(\tau)$ to show the intermittent properties in and near the MR region. The PDFs and kurtosis are deduced by calculating the field increments and analyzing the high-order moments of |B|. We define $\Delta B = (\delta B(\tau) - <\delta B(\tau)>)/\sigma_B$, where $\delta B(\tau) = |B(t+\tau) - B(t)|$, $\sigma_B$ is the standard deviation of $\delta B$, $\tau$ is the time lag, and the angle brackets denote ensemble averaging over time. Then, the kurtosis is deduced by $\kappa(\tau) = <|\Delta B|^4>/<|\Delta B|^2>^2 - 3$. For intermittent turbulence, the departure from the Gaussian distribution of the PDFs implies intermittency and, similarly, a Gaussian distribution would make the kurtosis equal to zero. Three different intervals are selected to represent the MR region (11:54-11:56UT) and the region far away from the MR event at both sides (11:46-11:48UT and 12:12-12:14UT). In Fig.3(a) and Fig.3(b), the homogenous structures revealed by the near-Gaussian PDFs and near-zero kurtosis are found in the region far the MR event. Meanwhile, the PDFs in the middle column reveals obvious departure from a Gaussian distribution and the kurtosis decreases from ~5 to ~1 with the time lag increasing from 0.18s to 0.64s. These phenomena indicate that obvious intermittent structures exist in the MR region at ion dissipation scales, which is also consistent with previous reports (Karimabadi, et al. 2013b; Osman, et al. 2011; Osman, et al. 2012a; Osman, et al. 2012b; Wan, et al. 2012).

We then analyze the fractal differences in these regions. The fractal nature of the turbulent fluctuations can be studied by comparing the scaling exponents deduced from the *p*th-order magnetic structure function $S(p,\tau)=<|\delta B(\tau)|^p>\propto \tau^{\varepsilon(p)}$. In a system with finite size, we can use the extended self-similarity (ESS) (Benzi et al. 1993) approach to improve the calculation by assuming $S(p,\tau) \propto S(3,\tau)^{\zeta(p)}$, where the exponents are normalized by $\zeta(p)=\varepsilon(p)/\varepsilon(3)$. If the deduced $\zeta(p)$ is linear in *p*, then it indicates monoscaling, while a nonlinear relation with *p* denotes that it is multifractal. Fig.3(c) shows the calculated scaling exponents from $\tau$=0.18s to 0.64s in each region. The obviously nonlinear relation in the center MR region and the linear behavior in the region far from the MR site apparently represent the multifractal and monofractal processes, respectively. Moreover, the scaling exponents in the MR region are consistent with a modified She-Leveque model (Müller & Biskamp 2000; Müller, Biskamp, & Grappin 2003) $\zeta(p)=p/g^2+1-(1/g)^{p/g}$ where *g*=3.0. In particular, the parameter *g*=3.0 indicates that the scaling property corresponds to the isotropic MHD intermittency model (Müller & Biskamp 2000; Müller, et al. 2003), and this result could also be compared to the recent observations (Kiyani et al. 2013). However, before more elaborate investigation are carried out, such as precisely decomposing the fluctuations into parallel and perpendicular directions (Kiyani, et al. 2013), it should be very cautious when drawing a definite conclusion for the anisotropic problem due to the currently limited sample points. Nevertheless, the above analyses provide the first observational evidence for the intermittent multifractal dissipation region scaling around an MR site.

Actually, our analyses indicate that the multifractal dissipation phenomenon is closely related to the MR processes. As seen in Fig.4, the scaling exponents $\zeta(p)$ vs. *p* are plotted and superposed as a function of time. To reveal the characteristics of the scaling more clearly, each $\zeta(p)$ is calculated with 2, 4, and 10 minute intervals and fit by using the structure functions from $\tau$=0.18s to 0.64s. In the 2 minute interval, it is found that, except in the center MR region at 11:54UT and 11:55UT, $\zeta(p)$ reveals a linear relation with *p* in other regions as the distance increases from the reconnection site. However, as the interval increases to 4 minutes, the nonlinear property of $\zeta(p)$ in the center MR region becomes less apparent, and $\zeta(p)$ finally shows a linear relation with *p* when the investigated interval extends to 10 minutes. These contrasting results indicate that the multifractal scaling effect introduced by MR would decay fast and mix with the background solar wind as it extends out from the reconnection site. If we regard the range 11:54-11:55UT as being dominated by the MR process, the dominant scale of this MR region could be roughly estimated to be ~500$d_i$. Moreover, aside from the regions referred to near the BL, we have also checked much more solar wind data provided around 1 AU. In most cases, the deduced $\zeta(p)$ in the ion dissipation range reveals a good linear relation with *p*, which is the same as the previous report (Kiyani, et al. 2009). Therefore, these results imply that the multifractal dissipation region scaling near the MR site is unique in the solar wind. It is quite different from the background solar wind turbulence and can only be observed near the MR region.

**Discussion and conclusion**
In the scaling computations, if the sample intervals are relatively short, then the *p*th-order magnetic structure function should not be empirically trusted when *p* is larger than 4 (Dudok de Wit 2004). To a finite length time series, it should be noted that the extremal values would strongly influence the scaling properties if the probability distribution is heavy tailed (Chapman et al. 2005; Kiyani, Chapman, & Hnat 2006). Therefore, we have removed 1% of the largest data points in $\Delta B$ before computing the scaling (in Fig. 3 and Fig. 4). Actually, removing 1% of the data is an empirical estimate. To ensure that the multifractal scaling is robust, it is necessary to examine the convergence properties of the original scaling exponents $\varepsilon(p)$ in the center MR region (11:54UT-11:56UT) by removing different amounts of data (Kiyani et al. 2007). Fig.5 shows the contrasting behaviors of $\varepsilon(p)$ after removing different outliers (0%-5%). $\varepsilon(p)$ is also deduce from the log-log fitting to $S(p,\tau) \propto \tau^{\varepsilon(p)}$ using the same fitting range as in previous calculations. After applying the ESS approach, $S(p,\tau)$ vs. $S(3,\tau)$ has a better linear property than $S(p,\tau)$ vs. $\tau$ in the log-log plot. So we can see that the error bars of $\zeta(p)$ in Fig.3 (c) are smaller than those of $\varepsilon(p)$ in Fig. (5). It is also shown in Fig.5 (a) that $\varepsilon(p)$ shows nonlinearity features similar to $\zeta(p)$ and it begins to approach linearity as the data points are successively removed. It is important to note that this convergence process procedes gradually, not suddenly. As seen in Fig.5 (b), $\varepsilon(2)$ displays a persistent secular drift with the continued removal of data instead of forming a plateau-like shape. Such properties (Kiyani, et al. 2006; Kiyani, et al. 2007) demonstrate evident multifractal scaling in the center MR region.

It should be noted that the MR region is relatively small and the MR-generated turbulence cannot be detected far away from the reconnection site. Hence, it is the crucial to choose short time intervals for the scaling analyses. To this end, as seen in Fig. 4, if longer intervals (10 minutes) are chosen, then the calculated scaling exponents in the center MR region would behave the same as in the nearby solar wind. Variations of the magnetic field in this region are totally smeared out and the involved MR processes could not be properly revealed. On the other hand, it seems unnecessary to choose even shorter intervals (e.g. 1 minute), since the multifractal scaling property has already been clearly revealed by using 2 minute intervals. Therefore, investigations in the 2 minute intervals are a compromise choice which could both guarantee the resolution and suppress the statistical uncertainty to a reasonable extent.

In conclusion, the above analyses show that the region near the MR site is more intermittent in the ion dissipation range than that in the 'background' solar wind turbulence. Most importantly, our results provide the first observational evidence that the region near the MR site follows a multifractal scaling law extending to the ion dissipation scale, which is quite distinct from the monoscaling processes found in the solar wind. This fractal difference is remarkable since it tends to imply that the dominant physical mechanisms in these different regions are not the same. In the MR region, the dominant mechanisms are MR processes during which the dissipation of magnetic energy has been demonstrated to proceed via a spatial multifractal field of structures generated by an intermittent cascade on kinetic scales (Leonardis, et al. 2013). Meanwhile, in the solar wind, if MR is also regarded as the main dissipation process, then multifractal scaling in the dissipation range would be expected, but such a phenomenon has not been found. Since MR occurs in small regions that do not fill the majority of the space, a measurement of the scaling of the structure functions of solar wind dissipation will most often not include a majority of the current sheets, and so will not indicate a multifractal scaling. Thus, previous analyses could all have missed the current sheet signature, and hence would not show multifractal scaling. If so, now is probably not the best time to answer whether MR is the main dissipation process in the solar wind. What fraction of the energy is dissipated by MR in the solar wind? What fraction is dissipated by wave particle interactions? How do their proportions change with the local solar wind conditions? To reach more definitive conclusions, it is necessary to check much more solar wind intervals containing MR with high temporal resolution data and to compare these results with numerical simulations. While current analyses clearly demonstrate that the region around an MR site reveals a unique multifractal scaling in the dissipation range, which is different from those in the solar wind, and this conclusion could be a stepping stone toward the resolution of the debate concerning the different dissipation mechanisms.

**Acknowledgements**

This work is jointly supported by the National Natural Science Foundation of China (41374174, 41231068, 41204123, and 41304131), the Specialized Research Fund for State Key Laboratories, the Open Research Program for Key Laboratory of Geospace Environment CAS, and the Science and Technology Development Fund of Macao SAR (039/2013/A2). We thank NASA CDAWEB for providing the Wind data. Y.W. thanks Dr. K. Kiyani, Dr. M. L.

# Figures

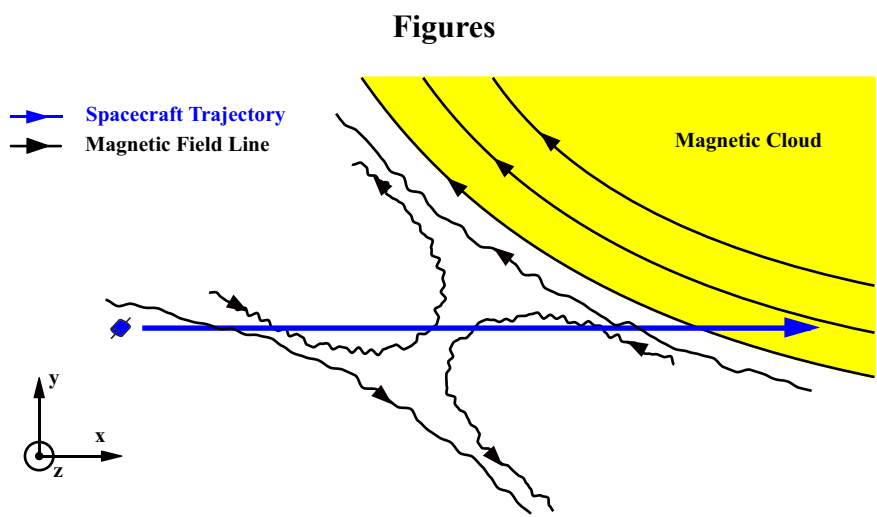

Fig 1: Illustration of a BL and the circumstances when a spacecraft encounters a BL. The geometric configuration of the MR region encountered by the Wind spacecraft between 11:30-12:30 UT on 1997 April 21 is also sketched.

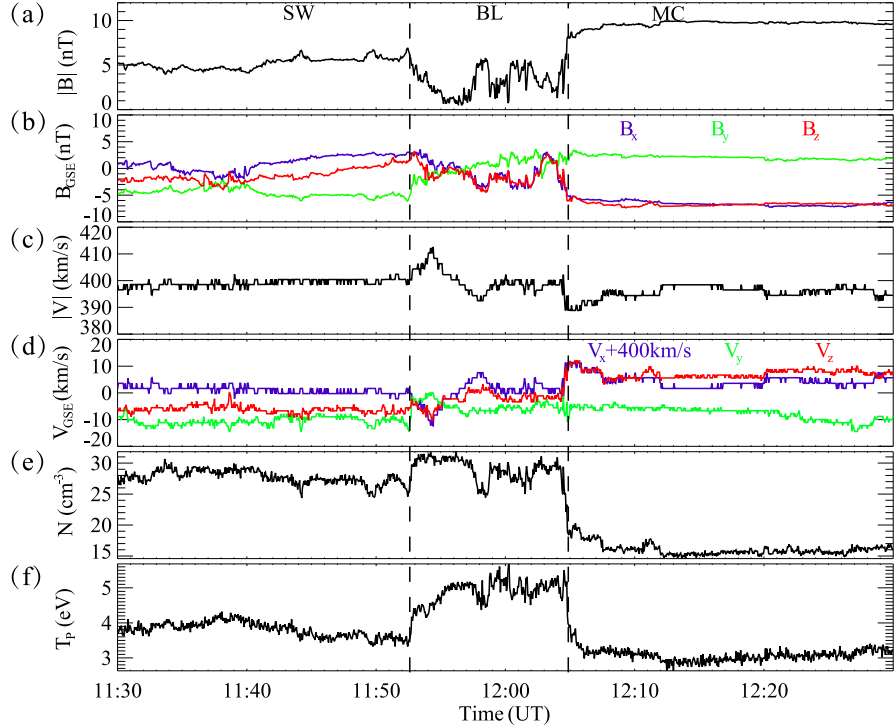

Fig 2: Detailed information measured by the Wind spacecraft between 11:30-12:30 UT on April 21, 1997. (a)-(b) The magnitude of the magnetic field and its components with a cadence of 0.092 seconds. (c)-(d) The velocity and its components with a cadence of 3 seconds (Vx has been shifted by 400 km/s). (e)-(f) Proton density and temperature with a cadence of 3 seconds. The dashed lines mark the boundary of the BL.

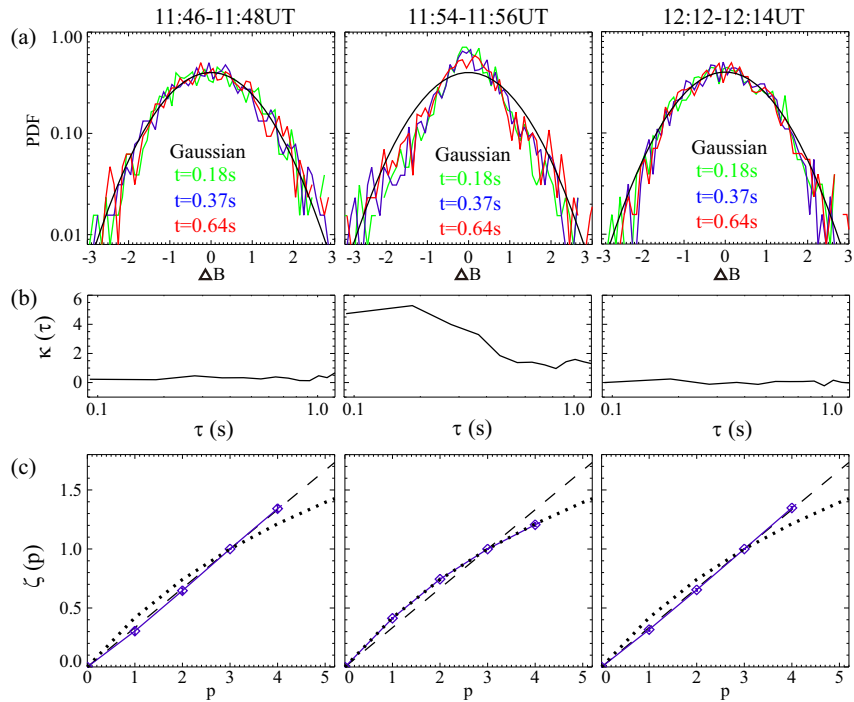

**Fig 3**: (a) PDFs of the |B| in different regions with different time lag. (b) Kurtosis as a function of the time lag in different regions. (c) Blue diamond denotes the scaling exponents $\zeta(p)$ calculated by the 4th order magnetic structure function. Black dashed line denotes $\zeta(p)=p/3$. Black dotted line denotes the modified She-Leveque model, $\zeta(p)=p/g+1-(1/g)^{p/g}$ with $g=3.0$. Error bars represent one standard deviation of uncertainty.

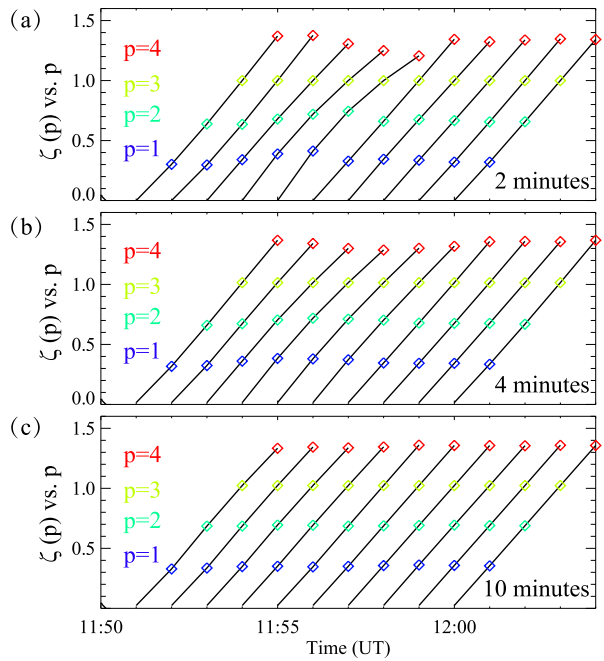

**Fig 4:** Scaling exponents ζ(*p*) vs. *p* superposed as a function of time deduced from 2 minute (a), 4 minute (b) and 10 minute (c) intervals. Different colors denote different moments *p*.

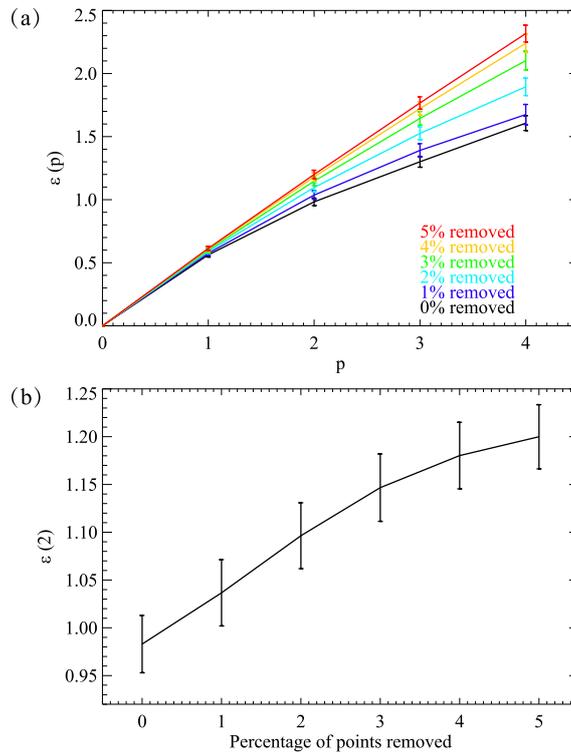

**Fig 5**: (a) Original scaling exponents $\varepsilon(p)$ vs. *p*. Different colors denote different percentages of points removed before computing. (b) $\varepsilon(2)$ vs. percentage of points removed. Error bars represent one standard deviation of uncertainty.